\documentclass[apjl]{emulateapj}

\newcommand{\kms}{\,km\,s$^{-1}$}       \newcommand{\sqcm}{\,cm$^{-2}$}  
\newcommand{\hi}{\ion{H}{1}}
\newcommand{\tm}{\tablenotemark}       \newcommand{\tn}{\tablenotetext}
\newcommand{\hst}{\emph{HST}}          
\newcommand{\cf}{\ion{C}{4}}           \newcommand{\sif}{\ion{Si}{4}}
\newcommand{\siw}{\ion{Si}{2}}         \newcommand{\sit}{\ion{Si}{3}}
\newcommand{\cw}{\ion{C}{2}}           
\newcommand{\pds}{\object{PDS\,456}}   \newcommand{\nf}{\ion{N}{5}}

\begin{document}
\shorttitle{The Biconical Galactic Nuclear Outflow}
\shortauthors{Fox et al.}
\title{Probing the Fermi Bubbles in Ultraviolet Absorption:\\ 
A Spectroscopic Signature of the Milky Way's Biconical Nuclear 
Outflow\footnotemark[1]}
\footnotetext[1]{Based on observations taken under program 13448 
  of the NASA/ESA Hubble Space Telescope, obtained at the 
  Space Telescope Science Institute, which is operated by the Association of 
  Universities for Research in Astronomy, Inc., under NASA contract 
  NAS 5-26555, and under program 14B-299 of the NRAO Green Bank Telescope,
  which is a facility of the National Science Foundation operated under 
  cooperative agreement by Associated Universities, Inc.} 
\author{Andrew J. Fox$^{2}$,  
Rongmon Bordoloi$^2$, Blair D. Savage$^3$, 
Felix J. Lockman$^4$, Edward B. Jenkins$^5$, 
Bart P. Wakker$^3$, Joss Bland-Hawthorn$^6$, 
Svea Hernandez$^2$, Tae-Sun Kim$^7$,
Robert A. Benjamin$^8$, David V. Bowen$^5$, \&
Jason Tumlinson$^2$} 
\affil{$^2$ Space Telescope Science Institute, 3700 San Martin Drive,
  Baltimore, MD 21218\\
$^3$ Department of Astronomy, University of
 Wisconsin--Madison, 475 North Charter St., Madison, WI 53706\\
$^4$ National Radio Astronomy Observatory, P.O. Box 2, Rt. 28/92, 
Green Bank, WV 24944\\
$^5$ Princeton University Observatory, Princeton, NJ 08544\\
$^6$ Institute of Astronomy, School of Physics, University of Sydney, 
NSW 2006, Australia\\
$^7$ Osservatorio Astronomico di Trieste, Via G.B. Tiepolo 11, 
34143 Trieste, Italy\\
$^8$ Department of Physics, University of Wisconsin--Whitewater, 
800 W. Main St., Whitewater, WI 53190\\}
\email{afox@stsci.edu}

\begin{abstract} 
Giant lobes of plasma extend $\approx$55\degr\ above and 
below the Galactic Center, glowing in emission 
from gamma rays (the Fermi Bubbles) to 
microwaves (the WMAP haze) and polarized radio waves.
We use ultraviolet absorption-line spectra from the
\emph{Hubble Space Telescope} to constrain the velocity 
of the outflowing gas within these regions, targeting the quasar PDS\,456 
($\ell,b$=10.4\degr, +11.2\degr).
This sightline passes through a clear biconical structure seen 
in hard X-ray and gamma-ray emission near the base of the northern Fermi Bubble.
We report two high-velocity metal absorption components,
at $v_{\rm LSR}$=$-$235 and +250\kms, %*
which cannot be explained by co-rotating gas in the Galactic disk or halo.
%gas in the low halo of the inner Galaxy co-rotating with the disk
Their velocities are suggestive of an origin on the 
front and back side of an expanding biconical outflow emanating from 
the Galactic Center. We develop simple kinematic biconical outflow models 
that can explain the 
observed profiles with an outflow velocity of $\ga$900\kms\ %* 
and a full opening angle of $\approx$110\degr\ (matching the X-ray bicone). %*
This indicates Galactic Center activity over the last $\approx$2.5--4.0\,Myr, 
in line with age estimates of the Fermi Bubbles.
The observations illustrate the use of UV spectroscopy to
probe the properties of swept-up gas venting into the Fermi Bubbles.
\end{abstract}

\section{Introduction}
The nuclei of star-forming galaxies are the
powerhouses where super-massive black holes collect interstellar gas 
from their surroundings, funnel it onto accretion disks, and heat it to 
extreme temperatures. 
The energy released from these environments and from supernovae 
drives gas out of galaxies through large-scale Galactic winds
\citep[see reviews by][]{He02, Ve05}.
Many outflows in nearby star-forming galaxies are nuclear
in origin and biconical in shape, as in
\object{NGC\,3079} \citep{Ce01, Ce02}, 
\object{M82} \citep{BT88, SB98, Oh02}, and 
\object{NGC\,1482} \citep{VR02}.

Our vantage point inside the dusty rotating disk of the Milky Way
hampers our knowledge of any \emph{Galactic} nuclear outflow. 
\citet{Lo84} noted an absence of \hi\ in the inner Galaxy and suggested it
had been cleared out by a wind. 
The first clear detection of a biconical outflow was made in mid-IR emission 
and (on larger scales) in hard X-ray emission \citep{BH03}.
It has since become clear that
the Galactic Center (GC) lies between two giant, 
energetic lobes associated with outflowing gas. These lobes extend 
$\approx$55\degr\ above and below the GC ($\approx$12\,kpc) 
and show enhanced emission across the 
electromagnetic spectrum, including:
(1) $\gamma$-ray emission, i.e. the Fermi Bubbles 
\citep[FBs;][]{Su10, Do10, Ac14}; 
(2) soft X-ray emission \citep[0.3--1.0\,keV;][]{Sn97, Ka13}; 
(3) K-band microwave emission (23--94\,GHz), known as the so-called 
WMAP haze \citep{Fi04, DF08}; %*
(4) polarized radio emission at 2.3\,GHz 
\citep[synchrotron radiation;][]{Ca13}. 
Near the base of the FBs, within 700\,pc of the GC, 
\citet[][hereafter MG13]{MG13} recently discovered a population of 
$\approx$100 small \hi\ clouds whose kinematics
are consistent with a biconical outflow with a velocity of $\approx$200\kms. 
%They found such clumps imply a hot wind velocity of $\approx$600--1900\kms.

\begin{figure*}[!ht]
\epsscale{1.0}\plotone{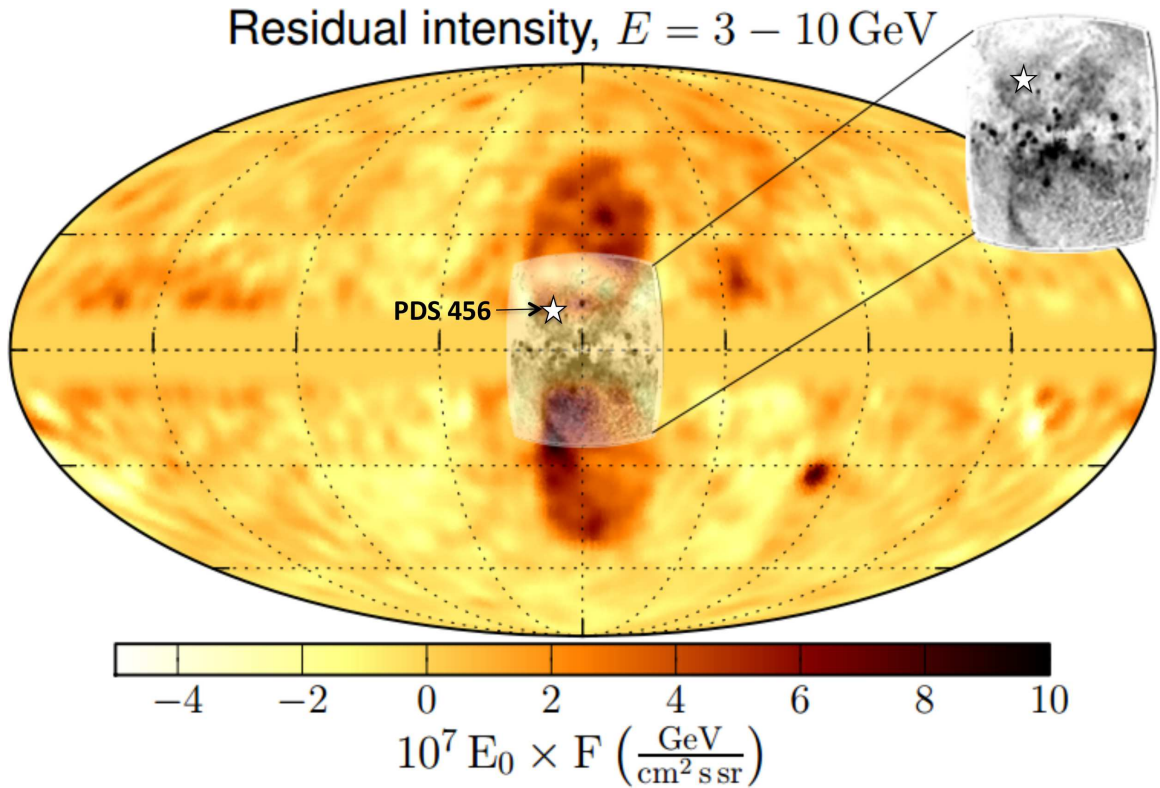}
\caption{Collage of gamma-ray and X-ray emission showing the striking biconical 
nuclear structure intercepted by the PDS\,456 sightline.
The yellow/orange map is an all-sky {\it Fermi} image of the residual gamma-ray 
intensity in the 3--10\,GeV range, in Galactic coordinates centered on the GC 
(Ackermann et al. 2014; \copyright\ AAS. Reproduced with permission). 
The Fermi Bubbles are the twin lobes in dark orange
at the center of the figure.
Superimposed in gray-scale is the {\it ROSAT} diffuse 1.5\,keV emission map,
based on Snowden et al. 1997, Bland-Hawthorn \& Cohen 2003, and 
Veilleux et al. 2005. The inset on the right shows a zoom-in on the
X-ray data.
Adapted from Figure 22, \emph{The Spectrum and Morphology of the Fermi 
Bubbles}, M. Ackermann et al., ApJ, Volume 793, Issue 1, 2014, Page 64.}
\end{figure*}

Clearly, a nuclear outflow is being driven out from the GC.
Yet only a handful of GC sightlines have published UV spectroscopy that
constrains the kinematics, ionization state, and elemental abundances
of the nuclear outflow, including two AGN directions \citep{Ke06} and 
several stellar directions \citep{Bo08, Ze08, Wa12}. %Zs03
However, none of these UV studies have probed a full sightline through the 
front and back sides of the FBs within 20\degr\ of the GC, where the 
$\gamma$-ray emission, X-ray emission, and (presumably) wind activity 
is strongest. The only such sightline with published data
is the \emph{optical} spectrum of the blue supergiant 
\object{LS\,4825} ($\ell,b$=1.7\degr, --6.6\degr, $d$=21$\pm$5\,kpc),
which shows complex multi-component \ion{Ca}{2} and \ion{Na}{1} profiles 
spanning $\approx$300\kms\ \citep{Ry97}.

We have initiated an observing program with the \emph{Hubble Space Telescope}
(\hst) to study the gas in the GC region 
(defined here as $0\degr\!<\!l\!<\!30\degr$ and $330\degr\!<\!l\!<\!360\degr$)
with UV spectroscopy (Program IDs 12936 and 13448).
The quasar \pds\ ($z_{em}$=0.184, $\ell,b$=10.4\degr,+11.2\degr, also known
as \object{IRAS\,17254-1413}) is
the lowest latitude and smallest impact parameter to the GC
($\rho$=2.3\,kpc) of any AGN in our sample.
Furthermore, this sightline is \emph{the only AGN direction in our sample 
that passes through the biconical region of enhanced ROSAT
1.5\,keV X-ray emission} centered on the GC \citep{Sn97}, where 
the $\gamma$-ray emission is also strong since the direction intersects
the base of the northern FB \citep{Su10}. 
The \pds\ direction (see Figure 1) is therefore of high interest for 
looking for UV outflow signatures.
There are no known 21\,cm (neutral) high-velocity clouds (HVCs) 
in this direction \citep[e.g.][]{Pu12}. % \citep{Wa01}

In this Letter we present new UV and radio spectra of \pds\
to explore the properties of gas entrained
in the Galactic nuclear outflow. 
In Section 2 we discuss the observations and their reduction. 
In Section 3 we present the UV absorption-line spectra and discuss the 
identification of outflow components. 
Motivated by the component structure observed in our spectra,
we present numerical kinematic models of a nuclear biconical outflow in 
Section 4. In Section 5 we present a discussion of our results.
A full discussion of other directions that penetrate
the northern and southern FBs will be presented
in an upcoming paper.

\section{Observations and Data Reduction}

\subsection{COS Spectra}

\pds\ was observed on February 10 2014 with the Cosmic Origins Spectrograph
\citep[COS;][]{Gr12} onboard \hst\ for a total of five orbits. 
The observations used the G130M/1291 and G160M/1600 grating/central
wavelength settings, all four FP-POS positions,
and exposure times of 4846\,s for G130M and 8664\,s for G160M.
Individual exposures were aligned in velocity space using the centroids of 
known low-ion interstellar absorption lines,
and then co-added following the same procedures as described in \citet{Fo14}.
The spectra have a velocity resolution (FWHM) of $\approx$20\kms,
a signal-to-noise ratio near the absorption lines of interest 
of $\approx$12--20 (per resolution element), 
an absolute velocity scale uncertainty of $\approx$5\kms, %&
and cover the wavelength interval 1133--1778\,\AA, with
small gaps between detector segments at 1279--1288 and 1587--1598\,\AA.
The spectra were normalized around each absorption line using linear continua
and for display are rebinned by seven pixels (one resolution element),
though the Voigt-profile fits (described below) were made on the unbinned data.

\subsection{GBT Spectra}
We obtained several deep \hi\ 21\,cm pointings of the \pds\ direction
using the Green Bank Telescope (GBT) under NRAO program GBT/14B-299,
with the goal of detecting the HV components in emission.
Multiple scans of \pds\ were taken on October 3
and October 4 2014 with the VEGAS spectrometer
in frequency-switching mode, for a total of 35 minutes of integration.
The data were taken by frequency-switching either 3.6 or 4.0\,MHz, resulting
in an unconfused velocity range of at least 760\kms\ about zero velocity
at an intrinsic channel spacing of 0.151\kms.
The spectra were Hanning smoothed, then
calibrated and corrected for stray radiation using the procedure described
by \citet{Bo11}.  One of the receiver's two linear polarizations (PLNUM=1)
gave consistently superior instrumental baselines so only those data were
used.  A fourth-order polynomial was removed from emission-free portions of
the final average. The resulting spectrum has an rms brightness
temperature noise of 17.5\,mK in a 0.30\kms\ channel, giving a 1$\sigma$
sensitivity to a line 40\kms\ wide (chosen to be typical of Galactic HVCs)
of $N$(\hi)=1.1$\times$10$^{17}$\sqcm.

\section{Identification of Outflow Components}

\begin{figure}[!ht]
\epsscale{1.0}\plotone{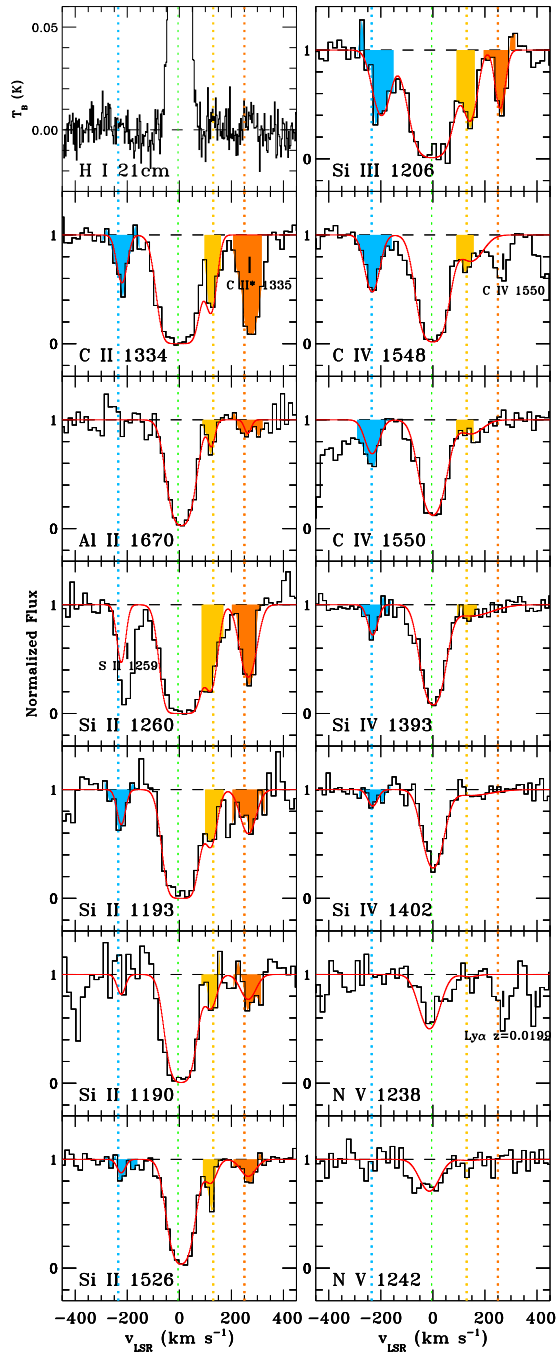}
\caption{\hst/COS and GBT spectra of the quasar PDS\,456.
Normalized flux is plotted against LSR 
velocity for all UV metal absorption lines that show high-velocity (HV)
absorption (with low ions in the left column and intermediate/high ions
in the right column). 
The GBT \hi\ 21\,cm emission spectrum is included in the 
top-left panel. Absorption-line components are observed at 
$v_{\rm LSR}$=$-$235, $-$5, +130, and +250\kms; only the $-$5\kms\ component
is seen in 21\,cm emission. Red lines indicate Voigt-profile fits.
The $-$5\kms\ (unshaded) and +130\kms\ (yellow) components
have velocities consistent with co-rotating foreground gas. However, the
$-$235\kms\ (blue) and +250\kms\ (orange) components %*
cannot be explained by co-rotation; instead, their velocities
are suggestive of gas swept up by a biconical outflow.
In this scenario the near-side of the outflowing cone gives the 
$-$235\kms\ component and the far-side gives the +250\kms\ component.} 
\end{figure}

\begin{figure}
\epsscale{1.1}\plotone{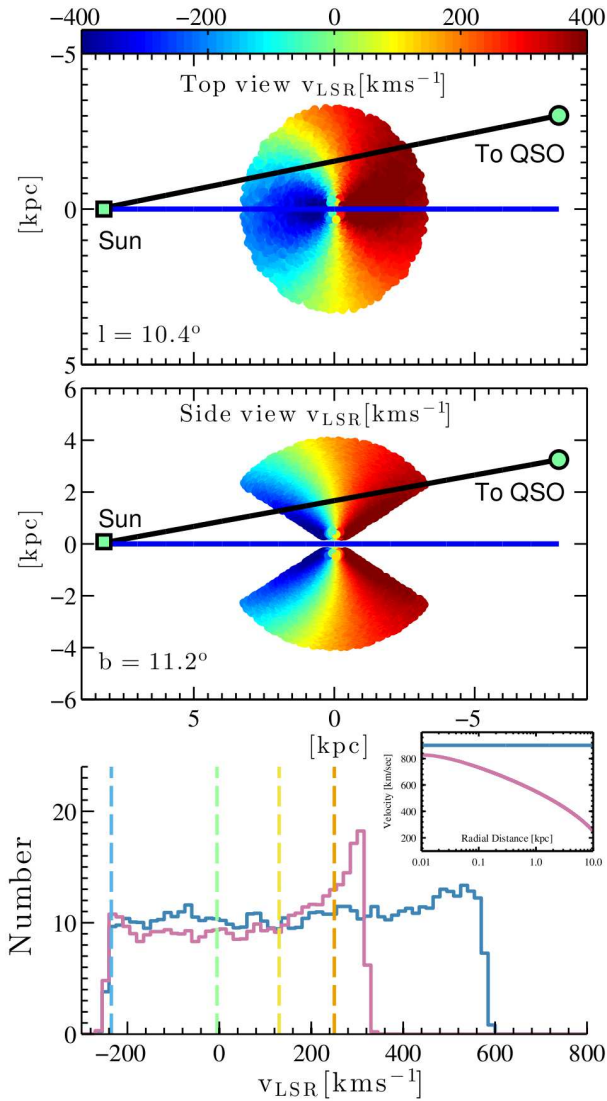}
\caption{Numerical models of the Galactic biconical nuclear
outflow, which can explain the observed absorption-line kinematics.
The models have a constant outflow velocity of 900\kms\ and a %*
full opening angle of 110\degr\ (tuned to match the X-ray bicone).
We investigate two sets of models: constant-velocity 
and momentum-driven. 
{\bf Top:} 
Top view of the momentum-driven outflow, looking down on the Galactic plane.
Each outflow particle is color-coded by its LSR velocity. The near side of 
the outflow is blueshifted and the far-side of the outflow is redshifted.
{\bf Middle:}
Side view of the momentum-driven outflow, showing the latitude where the 
PDS\,456 sightline intercepts the bicone.
{\bf Bottom:} Distribution of LSR velocities of the outflow particles 
along the PDS\,456 sightline in 100 realizations of the models,
for both the constant-velocity (blue line) and momentum-driven 
(purple line) cases. The inset panel shows the velocity profile of the two 
models.
The centroids of the observed components in the 
PDS\,456 spectrum are shown with vertical lines.}
\end{figure}

The COS spectra of \pds\ show four absorption components (see Figure 2), 
centered at $v_{\rm LSR}$=$-$235, $-$5, +130, and +250\kms. %*
Absorption is seen in low-ionization (\cw, \siw, \ion{Al}{2}), 
intermediate-ionization (\sit), and high-ionization (\cf, \sif, \nf) 
species, though the relative 
strength of absorption differs between components.
The $-$235, $-$5, and +130\kms\ components show absorption in the
low, intermediate, and high ions, whereas the +250\kms\ component 
is seen in the low- and intermediate ions only (no \cf, \sif, or \nf). %*
\hi\ emission in the GBT spectrum is only seen in the $-$5\kms\
component; in the other three components no \hi\ detection is made down to
a sensitive 3$\sigma$ upper limit $N$(\hi)$<$3.3$\times$10$^{17}$\sqcm.
The ionic column densities in the absorption components are given in 
Table 1. These were determined by fitting Voigt profiles to the data
with the {\tt VPFIT} software\footnote{Available at 
http://www.ast.cam.ac.uk/$\sim$rfc/vpfit.html}, using simultaneous
fits to all available lines of a given ion.

Foreground gas in the rotating disk of the Galaxy produces absorption at a 
range of LSR velocities, and a simple model of Galactic rotation 
can be used to predict the maximal allowed velocities for a
given latitude and longitude \citep[e.g.][]{SM87}, assuming 
cylindrical co-rotation.
Absorption detected \emph{outside} this velocity interval can 
be attributed to inflow or outflow. Toward \pds\ the minimum and maximum 
LSR velocities are $\approx$0\kms\ (at a distance of 0.0\,kpc) and 
+174\kms\ (at 8.8\,kpc), respectively. Therefore:\\
(1) the strong $-$5\kms\ component centered can be 
(either partially or completely) 
explained by foreground gas in the Galactic disk.\\
(2) the +130\kms\ component corresponds to distances of 7.1 and 9.9 kpc 
for co-rotation and so could also be tracing Galactic material
at $z$-distances of 1.4 and 1.9\,kpc, respectively. %7.1*sin(11.2 or 9.9)
However, for distant inner Galaxy sightlines, the assumption of 
co-rotation breaks down at $|z|\!>\!1$\,kpc \citep{Sa90, Se91},
potentially because the gas has been cleared out by a wind.
Thus the +130\kms\ component might trace a nuclear outflow,
but we cannot conclusively determine its origin.\\
(3) the $-$235 and +250\kms\ components are at least 
235 and 76\kms\ away from co-rotation, respectively, %*
and are therefore of most interest for a nuclear outflow. 
In the Galactic Standard of Rest (GSR), their velocities
are $-$190 and +295\kms, assuming the rotation velocity at the 
solar circle is +254\kms\ \citep{Re09}, and where 
$v_{\rm GSR}$=$v_{\rm LSR}$+254\,sin\,$\ell$\,cos\,$b$.
The velocities of these two components are %qualitatively and quantitatively 
consistent with an origin on the front (approaching) and back (receding) 
sides of a biconical nuclear outflow emanating from the GC, a scenario 
modeled in Section 4 and discussed in Section 5.

We note that the observed low-velocity high-ion absorption
(\cf, \sif, and \nf) is strong compared to other (non-GC) Galactic halo 
sight lines \citep{Sa01, Wa12}. In particular, \cf\ shows a 
$\approx$0.5\,dex excess in low-velocity column density,
%<log N(C IV)>=14.17, PDS456: 14.71 in Svea's VPFIT
which may be another sign of GC activity.

\section{Biconical Outflow Numerical Models}
We develop simple numerical models to explore the kinematic predictions
of a biconical nuclear outflow. The models are based on  
the \ion{Mg}{2} outflow models of \citet{Bo14}. 
We assume the outflow is an non-rotating expanding bicone centered on 
the GC with a constant mass flux. A population of $10^7$ test particles
is inserted at the base of the outflow, with a uniform filling factor.
The three principal free parameters in the models are the initial
outflow velocity $v_{out}$, the full opening angle $\alpha$ of the bicone,
and the velocity profile.
We take $\alpha$=110\degr\ based on the bicone %*
seen in the \emph{ROSAT} 1.5\,keV image (Figure 1) and investigate 
the value of $v_{out}$ needed to explain the $-$235 and +250\kms\ components.
We explore two cases for the velocity profile: 
constant-velocity (the simplest case), and
momentum-driven, where the outflow climbs ballistically out of the 
Galactic potential after being given an initial impulse
(e.g. by ram pressure from a hot wind).
In the latter case we use the formalism of \citet{Mu05} and 
\citet{DK12} to express the velocity as a function of radius from the GC.
We assume the outflow reaches a radial distance of at least 4\,kpc, distant 
enough to ensure the \pds\ sightline fully intercepts the bicone 
(see Figure 3, middle panel). 
The models are initially computed in the GSR reference frame, but are
transformed into the LSR frame for comparison with the data.
For both the constant-velocity and momentum-driven cases,
100 realizations of the models are created, and for each one
a line-of-sight with the \pds\ coordinates was generated and the kinematic
structure measured. The models do not account for the physical or kinematic 
properties of the external gaseous medium that confines the flow. 

The models are shown in Figure 3. We find $v_{\rm out}\!\ga\!900$\kms; 
lower velocities do not account for the $-$235\kms\ component.
Figures 3a (top view) and 3b (side view) show the velocity structure of the 
outflow as observed from vantage points outside the Galaxy. 
These plots illustrate the correspondence between LSR velocity and 
distance along the line-of-sight. The velocity fields are asymmetric
with respect to the vertical because of the transformation from the 
GSR to LSR frame, i.e. because we are observing the outflow from a 
co-rotating frame at a fixed distance.
Figure 3c shows the distribution of outflow velocities projected onto the 
\pds\ line-of-sight drawn from 100 realizations of each of the two models.
Discrete kinematic structure is seen in any single realization of the model 
(not shown in figure), but the component velocities differ substantially
between any two model runs, so we show the distribution to indicate
the \emph{range} of velocities predicted.

There are several points to note from the model velocity distributions.
First, the distributions are fairly flat for both constant-velocity 
and momentum-driven winds, which reflects our choice of model 
in which gas exists with uniform filling factor within the outflow, 
not just at the edges.
The prediction of outflowing gas with LSR velocities
near 0\kms\ could explain the 0.5\,dex excess in 
low-velocity high-ion absorption seen toward \pds.
Second, the distributions are \emph{asymmetric} around zero, ranging from 
$\approx$--250\kms\ (this velocity is a consequence of the requirement that we 
explain the $-$235\kms\ component toward \pds) to $\approx$+550\kms\ 
for the constant-velocity wind, and $\approx$+300\kms\ for the
momentum-driven wind.
This asymmetry is a projection effect arising because of the finite distance 
between the Sun and the GC (the near-field effect). 
Third, the momentum-driven wind model is more successful in
reproducing the range of velocity components seen toward \pds\
than the constant-velocity wind model, since the latter predicts
gas in the range $\approx$300--600\kms, which is not observed.
Models in which gas flows out preferentially along the edges of the
cone are also consistent with the data, but are outside the scope
of this Letter.
In summary, the simple momentum-driven biconical wind
model is able to reproduce the velocities of the HV components toward \pds.

\section{Discussion}
The \hst/COS spectrum of \pds, a QSO lying only 15.2\degr\ from the GC
in a direction of enhanced X-ray and $\gamma$-ray emission intercepting
the base of the northern FB, shows UV metal-absorption-line components at 
$v_{\rm LSR}$=$-$235 and +250\kms\ (corresponding to 
$v_{\rm GSR}$=--190 and +295\kms), %*
velocities which cannot be explained by 
gas in the low halo of the inner Galaxy co-rotating with the disk.
%gas in the co-rotating disk or halo of the Galaxy.
A further component at +130\kms\ is also difficult to explain via co-rotation.
None of the components shows 21\,cm emission in our GBT spectrum down to 
sensitive levels of $N$(\hi)$<$3.3$\times$10$^{17}$\sqcm\ (3$\sigma$),
which indicates the hydrogen is mostly ionized. %*
The kinematics of these components can be explained
in a scenario where cool swept-up gas is entrained on the near- 
(blueshifted) and far- (redshifted) side of a biconical outflow from 
the GC. 

There are several arguments supporting the biconical outflow explanation: 
(1) the \emph{ROSAT} X-ray imaging and \emph{Fermi} $\gamma$-ray imaging
both clearly indicate the presence of a biconical structure centered on the GC
and intersected by the \pds\ sightline;
(2) simple kinematic models of a biconical outflow (Section 4) naturally 
reproduce the presence of both negative- and positive-velocity gas components;
(3) such models also explain the excess low-velocity high-ion absorption;
(4) the MG13 results demonstrate a population of \hi\ clumps existing
closer to the GC with kinematics consistent with a biconical outflow;
(5) although we cannot rule out the possibility that individual 
HV absorbers arise in unrelated foreground or background HVCs,
which have a sky covering fraction in UV metal lines of $\approx$68--80\%
\citep{Co09, Sh09, Le12}, a \emph{chance alignment} of two unrelated HVCs 
at close-to-symmetric velocities of $-$235 and +250\kms\ 
would be needed to emulate the profiles of a biconical outflow;
(6) given the low latitude of the \pds\ sightline ($b$=+11.2\degr),
explaining the $-$235 and +250\kms\ components as being regular (non-GC) 
HVCs would require unusually large ($\ga$1000\kms) vertical inflow or 
outflow velocities, because of the projection factor sin $b$. This makes 
it highly unlikely that the \pds\ HV components are due to regular HVCs or
to tidally-stripped material like the Magellanic Stream, which have
much lower vertical flow velocities ($\sim$100--200\kms).

In the biconical outflow interpretation, the hot wind phase
is feeding the FBs with new plasma \citep[e.g.][]{Cr12}, 
whereas the cool low-ion gas represents
swept-up material entrained in the outflow. 
We stress that our \hst/COS observations do not directly probe the hot
phase, but instead probe cool ($T\!\sim\!10^4$\,K) entrained gas via 
the low ions and transition-temperature ($T\!\sim\!10^5$\,K) gas via 
the high ions,
which may trace the boundaries between the cool gas and the hot wind.
The cool outflowing gas has already been detected, albeit closer to the GC
with a much lower wind velocity, via the MG13 population of \hi\ clouds.
%It is possible the clouds are being ionized as they are driven out and 
%accelerated.
%These authors derive a velocity for the \emph{hot} wind phase of 
%$\approx$600--1900. %, in line with what we derive here.
Despite the disruptive instabilities that can destroy cold filaments and
clouds on short timescales, simulations show that such structures
can survive in the hot wind fluid 
if stabilized by magnetic fields or other mechanisms \citep{Co08, Mc14}.

Our models find the outflow velocity is $\ga$900\kms; 
such a flow must be $\approx$2.5--4.0\,Myr old to reach
2.3\,kpc, the impact parameter of the \pds\ sightline.
%IDL> print, (2.3*3.09d21)/(900.*1.e5*3.16e7)
This wind age is consistent with the timescale of energy injection
that created the FBs, whether via AGN jets \citep{GM12, Ya12}, feedback 
from nuclear star formation \citep{CA11, Ca13, La14}, accretion flows onto 
Sgr\,A$^*$ \citep{Ch11, Mo14}, or a spherical outflow from Sgr\,A$^*$ 
\citep{Zu11}.
It is also compatible with the observed H$\alpha$ emission from 
the Magellanic Stream, which is consistent with an 
origin following a Seyfert flare at the GC 1--3\,Myr ago \citep{BH13}.
%{\bf This is the first velocity-based age for GC activity.}
%Indeed, this timescale is consistent with AGN-driven and starburst-driven
%winds in many galaxies \citep{Ve05}.

In our upcoming work, we will analyze the UV spectra of 
other GC targets, including foreground stars at a range of distance
and background AGN at a range of latitude, both inside and outside
the FBs.
A large sample of sightlines is needed to fully characterize the 
extent and nature of the Galactic nuclear outflow.

\vspace {1cm}
{\it Acknowledgments.}
Support for program 13448 was provided by NASA 
through grants from the Space Telescope Science Institute, which is 
operated by the Association of Universities for Research 
in Astronomy, Inc., under NASA contract NAS~5-26555.
The National Radio Astronomy Observatory is a facility of the 
National Science Foundation operated under cooperative agreement 
by Associated Universities, Inc.
We thank Gerald Cecil for assistance with Figure 1
and the referee for an insightful report.
We thank Anne Franckowiak and the Fermi-LAT collaboration for permission to 
reproduce their all-sky residual intensity map.
JBH is funded by an ARC Laureate Fellowship.
TSK acknowledges support from a European Research Council Starting Grant 
in cosmology and the IGM under Grant Agreement (GA) 257670.

\begin{deluxetable*}{lccc cccc}
\tabletypesize{\small}
\tabcolsep=2.0pt
\tablewidth{0pt}
\tablecaption{Column Densities in the Metal Absorption Components 
toward PDS\,456\tm{a}}
\tablehead{$v_{\rm LSR}$ & log\,$N$(\siw) & log\,$N$(\sit) & log\,$N$(\sif) & log\,$N$(\cw) & log\,$N$(\cf) & log\,$N$(\nf) &
log\,$\frac{N({\rm C\,IV})}{N({\rm Si\,IV})}$\\
(km\,s$^{-1}$) & & & & & & &}
\startdata
--235      & 13.02$\pm$0.08 & 13.13$\pm$0.05 & 12.90$\pm$0.06 & 13.80$\pm$0.14 & 13.79$\pm$0.05 & ...            & 0.89$\pm$0.07\\
--5\tm{b}  & 14.76$\pm$0.05 & $>$14.04       & 14.05$\pm$0.05 & ...            & 14.71$\pm$0.05 & 14.09$\pm$0.05 & 0.66$\pm$0.05\\
+130\tm{b} & 13.40$\pm$0.05 & 13.06$\pm$0.05 & 13.03$\pm$0.16 & 14.14$\pm$0.12 & 13.58$\pm$0.06 & ...            & 0.28$\pm$0.17\\
+250       & 13.37$\pm$0.05 & 12.85$\pm$0.05 & ...            &            ... & ...            & ...            & ...
\enddata
\tn{a}{Errors are statistical only. No entry is given for blends, 
non-detections, or heavily saturated lines.}
\tn{b}{Velocity compatible with either a rotating disk or a nuclear outflow.}
\end{deluxetable*}

\end{document}